# Vibrotactile Signal Generation from Texture Images or Attributes using Generative Adversarial Network

Yusuke Ujitoko[1][0000-0001-6059-9324] and Yuki Ban[2, ✉][0000-0001-7349-6383]

[1] Hitachi, Ltd., Yokohama, Japan
yusuke.ujitoko.uz@hitachi.com
[2] The University of Tokyo, Chiba, Japan
ban@edu.k.u-tokyo.ac.jp

**Abstract.** Providing vibrotactile feedback that corresponds to the state of the virtual texture surfaces allows users to sense haptic properties of them. However, hand-tuning such vibrotactile stimuli for every state of the texture takes much time. Therefore, we propose a new approach to create models that realize the automatic vibrotactile generation from texture images or attributes. In this paper, we make the first attempt to generate the vibrotactile stimuli leveraging the power of deep generative adversarial training. Specifically, we use conditional generative adversarial networks (GANs) to achieve generation of vibration during moving a pen on the surface. The preliminary user study showed that users could not discriminate generated signals and genuine ones and users felt realism for generated signals. Thus our model could provide the appropriate vibration according to the texture images or the attributes of them. Our approach is applicable to any case where the users touch the various surfaces in a predefined way.

## 1 Introduction

The vibrotactile sense enables humans to perceive texture surface properties through tool-surface interaction. Unfortunately, the richness of the vibrotactile responses from virtual texture surfaces is missing from current tool-surface interactions on a touchscreen. The tool-surface interactions are composed of simple gestures such as tapping or flickering, so vibrotactile designer should find the appropriate vibrotactile sig-nals for each gesture. However, it is difficult to find ones. Though there are vibrotactile datasets which is made public, it is rare to find the appropriate vibrotactile signals from them. It is because such datasets contains at most 100 kinds of textures, as compared to countless kind of texture in the real world.

Instead of looking into datasets, vibrotactile modeling has been studied for a long time to provide such responses. However, there is no model that interactively generates vibrotactile responses based on the state of the tool and the state of the surfaces. Such model should learn the complex mapping between large input and output space; Inputs are state of the tool (ex. tool's velocity) and the state of the texture surface (ex.

texture's attributes), on the other hand, outputs are vibrotactile signals. Considering a limitation of a representational power that a trained single model can have, it is difficult to train the model that get both states of the tool and state of the texture surface as input. In other words, there is a trade-off between the model's interactivity for the tool's state and the one for the texture surface's state.

Emerging, recent data-driven approach for haptic modeling mainly focus on the interactivity of the tool's state. Prior studies mapped the normal force and the velocity magnitude of the tool with vibrational patterns [1, 2]. These vibrational patterns were encoded in the autoregressive model. Their model succeeded in mapping the tool's state and the vibration patterns. They are suitable for interactions where there is much variability with tool's velocity and applied force. However, the single model generating vibrational signals only supported single kind of texture that is used during training. Thus, when you try to generate vibrations of another kind of texture, you need to replace the model with another one.

This paper, on the other hand, focuses on the interactivity for the texture's state instead of the tool's state. Current touchscreen interactions are composed of simple gestures such as tapping or flickering, which is completed in a short time. With such gestures, the tool's velocity or applied force is approximately constant. On the other hand, such gestures are generally used for various texture surfaces. Thus, we pose the modeling task of generating appropriate vibrotactile signals that correspond to the visual information or attributes of texture. Such capabilities realize generating haptic signals for even unseen textures automatically or manipulating vibrotactile signals by changing attribute values. As an application of this model, we assume a vibrotactile designing toolkit for tool-surface interactions where designers can (1) set attributes of texture or (2) prepare texture images to generate the appropriate signals for gestures. The model that accomplishes it is required to have the capability to capture rich distribution.

Recently, generative methods that produce novel samples from high-dimensional data distributions, such as images, are finding widespread use. Specifically, Generative Adversarial Networks (GANs) [3] have shown promising results in synthesizing real-world images. Prior research demonstrated that GANs could effectively generate images conditioned on labels [4], texts [5], and so on. In spite of these promising results, there are few studies that used GANs to model time-series data distribution. Indeed, the generation of vibration by GANs has not been realized for now. In this study, we make full use of GANs for indirectly generating vibrotactile signals via time-frequency domain representation, which can be calculated as the image. We train the model so that it can generate vibrotactile signals conditioned on texture images or texture attributes.

The contribution of this study is three-fold. First, to our best knowledge, we introduce the problem of vibrotactile generation and are the first to use GANs to solve it. Second, we succeed in indirectly generate time-series data via time-frequency representation using GANs. Third, our trained single model meets the demand for interactiveness for the state of textures by providing the appropriate vibration that corresponds to the texture images or texture attributes.

## 2   Related Work

Modern data-driven texture modeling mapped the tool's state and the vibrational response of contact in autoregressive coefficients [1] or in neural network [2]. However, their single model could generate vibrotactile signals only for single texture. When generating vibrations of another kind of texture, the model was needed to be replaced with another one. Therefore, in this paper, we train the single model that generates appropriate haptic signals that correspond to the visual information or attributes of texture. The model that accomplishes it is required to have the capability to capture rich distribution.

Generative methods that produce novel samples from high-dimensional data distributions, such as images, are finding widespread use, for example in image-to-image translation [6], image super-resolution [7], and so on. In this study, we use GANs framework to generate sharp time-frequency samples for vibrotactile feedback. GANs were introduced in the seminal work of Goodfellow et al. [3], and are composed of two models: generator and discriminator. They are alternatively trained to compete with each other. Given a distribution, the generator is trained to generate samples from noise vector $z$ so that the generated samples resemble this true distribution. On the other hand, the discriminator is trained to distinguish whether the samples are genuine. After training, the generator can generate samples from noise vector $z$, which are indistinguishable from genuine samples by the discriminator.

Utilizing the GANs' capability to capture rich data distributions, there are several methods to manipulate the output samples of GANs. As one of the representative method, conditional sample generation has been studied. AC-GAN [8] is a promising variant of conditional GANs in which generator conditions the generated samples on its class label $c$ and the discriminator performs an auxiliary task of classifying the generated and the genuine samples into the respective class labels. In this setting, every generated sample is associated with a class label $c$ and a noise $z$, which are used by the generator to generate images $G(c, z)$.

By using conditional GANs, cross-modal generation has been studied. Text to image generation was implemented by [5]. In their studies, plausible images for birds and flowers were generated from their text descriptions. Cross-modal audio-visual generation was studied in [9]. Though they tried to generate spectrogram convertible to sound, they generated the rough spectrogram, which was not convertible to good sound. The reason for the poor generation is that the network architecture and optimization technique was not sophisticated. Inspired by their work, we construct the cross-modal visuo-vibrotactile generating model. We use the improved network architecture and optimization technique, and thus, our model is able to generate the spectrogram that is convertible to the vibrotactile signals.

## 3 Vibrotactile Signal Generation

### 3.1 Concept of Overall Model

By utilizing GANs' capability to capture rich data distributions, we would like to make the single generative model that has following features: automatic generation of vibrotactile signals either (1) from given texture images or (2) from given texture attributes. Though prior research focuses on the interactive generation based on tool's state, this paper proves the concept above for predefined tool's state under constrained touch interactions. Among various touch interactions, we focus on the task of moving a pen on the texture surface.

The overall diagram of our model is shown in Fig. 1. It consists of two parts: an encoder network, and a generator network. They are trained separately. The encoder is trained as an image classifier and it encodes texture images into a label vector $c$.

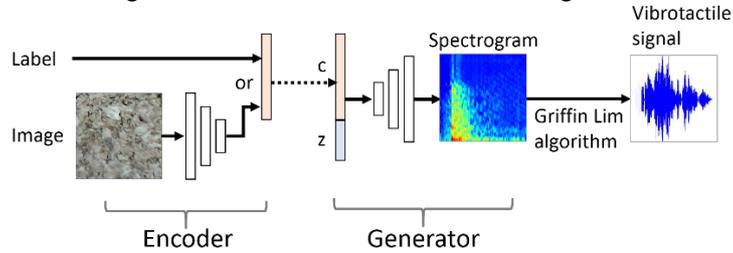

The generator is trained with discriminator in GANs training framework and generates spectrogram that is a representation of vibration in a time-frequency domain. We describe the training details for each network in the following sections. The overall model enables end-to-end generation from visual images or label attributes of texture to the vibrotactile wave.

**Fig. 1.** Overall diagram of our model.

We describe the data flow step by step based on Fig. 1. The input into the model is either a class label that represents the tactile attributes of the texture or a texture image. When the image is input, the label vector $c$ is extracted from the texture image through encoder network. The label vector $c$ is a categorical variable that shows the attributes of the texture. Next, the label vector $c$ is passed into the generator network. The generator concatenates the label $c$ and the random noise $z$ and transforms them into the spectrogram. The generated spectrogram is converted into the acceleration wave format by Griffin-Lim algorithm [10]. Then the wave format data is output to the user. With this overall model, users can input either label information or texture images to obtain vibration. That is why we do not adopt the network like pix2pix [6], which only supports input as images and converts images directory into signals.

Acceleration signals are used as vibrotactile stimulus in our model. In order to train the whole network, we use dataset [11], which contains acceleration signals and captured images during movement task. The pairs of signals and images are annotated with 108 classes.

## 3.2 Encoder

We trained the image encoder that encoded texture images into the label vector *c*. We adopted the deep residual network (ResNet-50) [12] architecture. We fine-tuned all the layers of the ResNet-50 that had been pre-trained with ImageNet [13]. We used Adam optimizer with a mini-batch size of 64. The learning rate started from 1e-3 and was decreased by a factor of 0.1 when the training error plateaued.

The size of provided images by [11] is 320 x 480. We fed them into the encoder network. For training phase of encoder network, we followed ordinary data augmentation settings. We scaled an image with factors in [1, 1.3], randomly cropped 128 x 128 size of it, flipped it horizontally and vertically, rotated it by a random angle. The recent data augmentation technique of random erasing and mixup were also used.

As a result of training, the trained encoder achieved a classification accuracy of more than 95 percent on the testing set. After the network was trained, its last layer was removed, and the feature vector of the second to the last layer having dimension of label vector was used as the image encoding in our generator network.

## 3.3 Generator

**Network Architecture and training settings.** Generator was trained with discriminator in GANs framework. During training, the discriminator learned to discriminate between genuine and generated samples, while the generator learned to fool the discriminator. Generator output samples $x = G(z, c)$ conditioned on both random noise vector *z* and a label vector *c* from dataset. Discriminator had two outputs: *D(x)* the probability of the sample *x* being genuine, and $P(x) = c$, the predicted label vector of *x*. After training, the discriminator was removed and the generator was only used in our model. Inspired by [14], we employed architecture and loss function, which was based on SRResNet[7], DRAGAN[15], and AC-GAN[8]. The architecture of generator and discriminator are shown in Fig. 2.

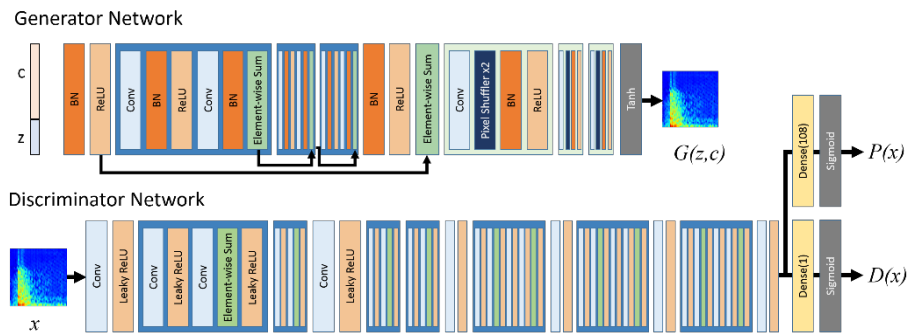

**Fig. 2.** Network architecture of generator and discriminator.

Acceleration signals orthogonal to the surface during movement task were used as vibrotactile stimulus and we aim at generating the signals by generator. For now, there are few studies generating time series data using GANs. It is because GANs are poor at generating time-series data though they are good at generating 2D images. Therefore, we chose amplitude spectrogram as a representation of the acceleration signals and trained GANs to generate spectrogram as if that was 2D image. The same dataset used for training encoder contained acceleration signals during movement task. Each signal had 4 seconds long and the sampling rate was 10 kHz. We computed the spectrogram from wave format using 512-point Short-Time Fourier Transform (STFT) with a 512 hamming window and a 128 hop size. Then, the linear amplitude of the spectrogram was converted to the logarithmic scale. We cropped the spectrogram and resized it into 128 x 128 size. As a result, the spectrogram contained the information of time-frequency domain up to 256 Hz for 1.625 seconds long. The values in the spectrogram were normalized into the range from 0 to 1.

We selected 9 textures out of 108 textures for GANs' training because it is stable to train conditional GANs with fewer number of conditional label dimensions. Thus, the dimension of categorical label $c$ was 9. On the other hand, the dimension of noise $z$ was 50. The selected 9 textures were representative of 9 groups of LMT haptic texture database [11] (Fig. 3). We used Adam optimizer with a mini-batch size of 64. The learning rate was fixed at 2e-4.

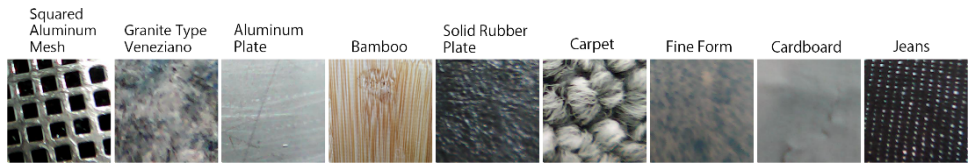

**Fig. 3.** Selected textures for GANs training.

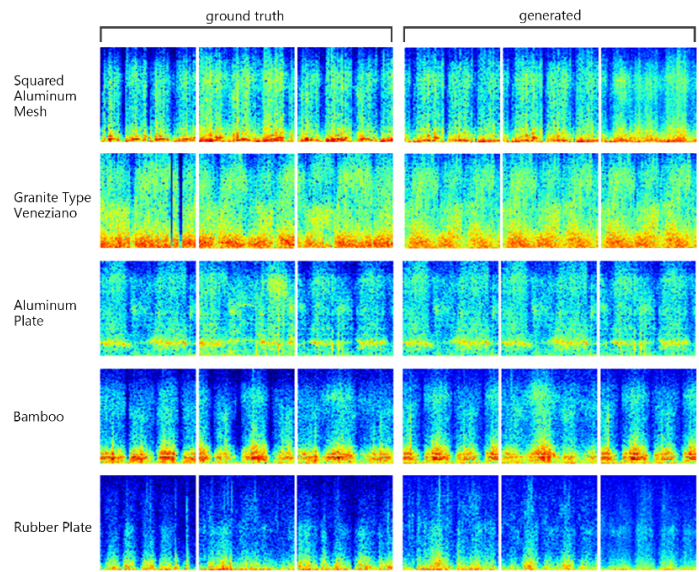

**Fig. 4.** The spectrogram for each class label in test dataset and the one generated by generator.

**Training Results of Generator.** The spectrogram in test dataset and the one generated by generator are shown in Fig. 4. The comparison between them shows the trained generator could generate the spectrograms that appear indistinguishable from test ones.

We describe the qualitative evaluation by user study in section 4. On the other hand, it is generally difficult to quantitatively evaluate the GANs. The ordinary evaluation metric of GANs, namely the "inception score" cannot to be applied to our case because the "inception score" is only applied to standard dataset such as CIFAR-10. Instead, we observe that t-SNE is a good tool to examine the distribution of generated images. A two dimensional t-SNE visualization is shown in Fig. 5. It is shown that the generated and test samples made group for each class label.

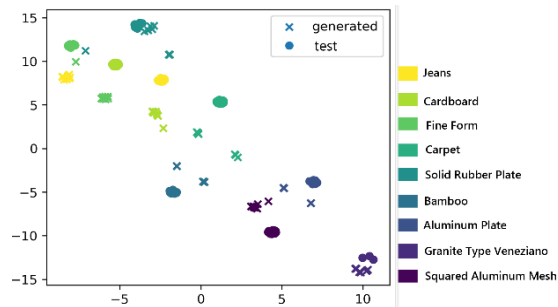

**Fig. 5.** t-SNE visualization of test dataset and the one generated by generator.

### 3.4 End-to end (E2E) Network

E2E cross modal generation of signals from texture images are realized by combining encoder and generator. The encoder was trained with 9 classes instead of 108 classes in accordance with input dimension of conditional generator. Fig. 7shows the generated spectrogram from the texture image and genuine one for each test image in the dataset. The comparison between them shows the E2E network could generate the spectrograms that seem indistinguishable from test one. We describe the qualitative evaluation by user study in section 4.

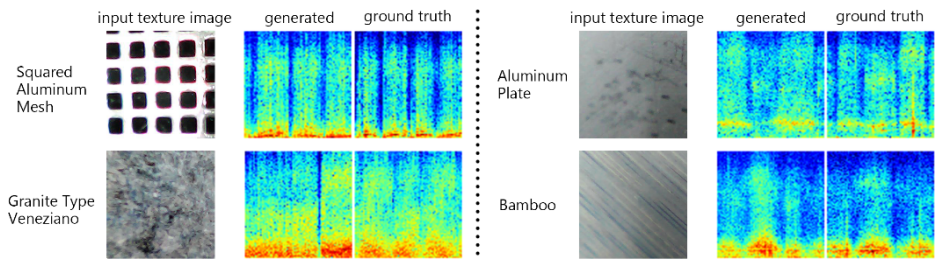

**Fig. 6.** E2E generation from texture images into spectrogram.

## 4   User study

User studies were conducted to investigate that whether our method could generate perceptually realistic vibrotactile stimuli. Two studies were conducted to evaluate Generator (Generator Ex.) and E2E network (E2E Ex.). Ten participants whose ages ranged from 22 to 25 (eight males and two females) participated in these studies. All of them were right handed. They were screened to determine that they were not depressed, terribly tired because the perception would be affected by physical or emotional states. Each of them participated in those two studies on another day. The data acquisition was approved by the University of Tokyo Ethics committee (approbation number: KE17-63) and written informed consent was obtained from all participants.

### 4.1   Experimental System

In user studies, participants' task was to move a pen-type device on a surface of a tablet device while receiving vibrotactile feedback. Our experimental system was constituted of the tablet device (Apple Inc., iPad Pro 9.7 inch), an amplifier (Lepai Inc., LP-2020A +), and a pen-type device with a vibrator (Fig. 8). The pen-type device, which we handcrafted, is specifically described in the next paragraph.

The pen-type device was about 20 g weight and about 140 mm long. The diameter of the grip part of the pen was about 10 mm. The pen tip wore conductive material that is ordinary used for the stylus. Since the shaft of the pen used in these studies was made of plastic and does not conduct to the grip part, we winded a conductive sheet on the grip to react with a capacitance type touch screen. Inside the pen-type device, the vibrator (ALPS Inc., HAPTIC™ Reactor) was embedded at the position of 2cm distance from the tip of the pen where participants gripped. The vibrator was small (35.0 mm × 5.0 mm × 7.5 mm) and light (about 5 g) enough not to prevent participants from moving the pen.

When participants touched and moved the pen on the surface, the vibration signal was output from earphone jack of the tablet, and amplified by the amplifier, and vibrator embedded on the pen presented the vibration to the participants' fingers

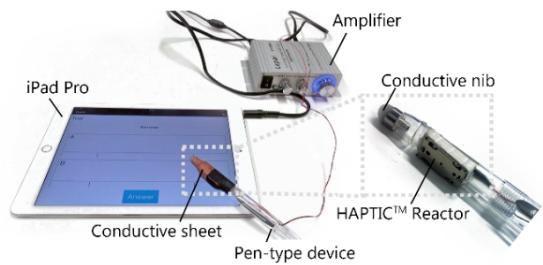

**Fig. 7.**   Setting of the experimental system.

### 4.2 Task Design

These studies used a within-participant design. Participants moved the pen-type device along the two different predefined path on screen in succession, while receiving either test or generated vibrational feedback. After that, participants tried to distinguish which stimulus was generated one. They also evaluated the realism of stimuli. In Generator Ex., generated signals were generated by feeding a label vector that represented each class into the generator. In E2E Ex., generated signals were generated by feeding a test image that represented each class into the encoder network. Corresponding class label texts or texture images were displayed on the touch screen. Participants' task was the same in Generator Ex. and E2E Ex. except that what they saw on screen was class label texts or texture images.

The procedure of one trial in participant's task is described in this paragraph. Participants moved the pen on a virtual texture surface from left to right for about 100 mm distance at fixed speed with their dominant hands. To control the movement speed and distance, the touch screen visualized a bar that indicated where and how much speed to move. According to the bar elongation, participants moved the pen approximately 100 mm distance in 1.6 seconds. Participants were told to hold the pen at the position where a vibrator was embedded. After completing movement on two surfaces, they answered which stimulus was felt generated one by tapping one of the two answer buttons. Besides, they answered the degree of realism for each stimulus by visual analogue scale (VAS) ratings [16] (Fig. 9 Right). Participants rated the realism that they felt on an analogue scale in this testing method. They answered the question "How much realism did you feel?" by rating realism on a 100 mm line on the touch screen anchored by "definitely not" on the left and "felt realism extremely" on the right. They used the pen-type device to check on this line. The displayed order of test and generated stimuli in one trial was shuffled.

Vibration signals that belonged to nine classes of textures that are modeled in Section 3 (Fig. 4) were prepared for this study. Test signals are randomly extracted from test dataset corresponds to each class, and generated signals are generated for each trial. Participants performed the trial ten times for each class. Therefore, each participant performed 180 trials in total for both Generator Ex. and in E2E Ex. E2E Ex. was conducted after Generator Ex. and these studies were held on separate days in order to prevent any satiation effects. To prevent sequential effects, the presentation order of these factors was randomly assigned and counter-balanced across participants.

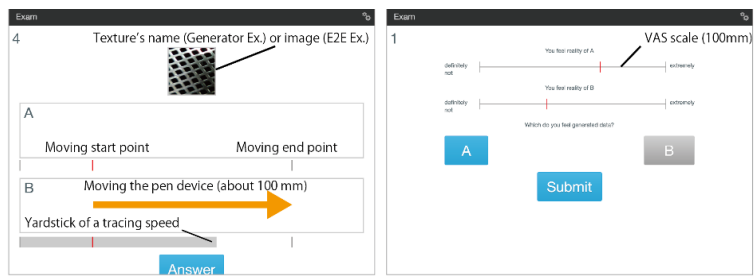

**Fig. 8.** Experimental windows.

### 4.3 Result

Fig. 10 shows the percentage of correctly identifying which stimulus was generated. We call this value as "Correct answer rate". If this value is close to 50 %, it means that participants failed to distinguish test data from generated data. Thus, it is confirmed that our method could generate the vibrational stimuli that were close to the genuine stimuli. The average and the standard error (SE) of "Correct answer rate" were 47.7 ± 1.49 % for Generator Ex., and 48.2 ± 2.49 % for E2E Ex. To investigate whether these rates were out of 50 %, we applied the Chi-Square goodness of fit test. It revealed that the rate of Carpet and Fine Foam condition in E2E Ex. were significantly lower than 50% (Carpet: $p<0.01$, Fine Foam: $p<0.05$).

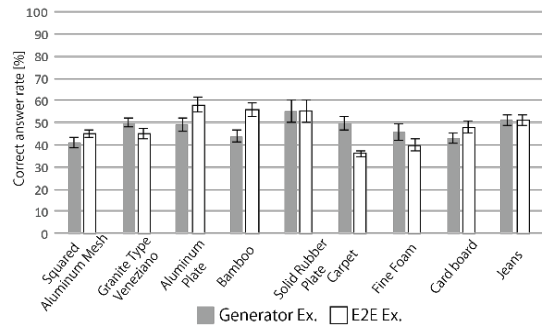

**Fig. 9.** Percentage of identifying which stimuli were generated one out of two.

On the other hand, Fig.11 shows the results of how much realism participants felt for each class. The score of the realism of test data was 72.9 ± 1.49 and that of generated data was 73.1 ± 2.93 in Generator Ex. In E2E Ex., the score of the realism of test data was 71.4 ± 2.04 and that of generated data was 70.3 ± 1.81. We used a Student's paired t-test to each texture condition, and revealed that there were significant differences between the average score of test and generated data for Bamboo in Generator Ex. ($p=0.025$), and Squared Aluminum Mesh, Bamboo, Card board in E2E Ex. ($p=0.026, 0.025, 0.025$). There was no significant difference between the average score of test and generated data in total.

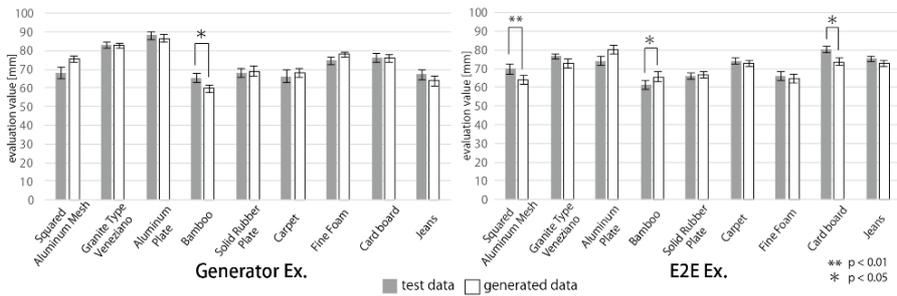

**Fig. 10.** The score of realism participants felt.

### 4.4 Discussion

"Correct answer rate" of most texture conditions were almost 50%, thus participants could not distinguish test data from generated data. In the post questionnaire, all participants answered that they did not find the difference between the test stimulus and the generated stimulus. Therefore, it can be said that our system has the potential to generate the high realistic vibrotactile signals from given texture attributes or given texture images. "Correct answer rate" of Carpet and Fine Foam class in E2E Ex. were significantly lower than 50% so that participants tended to misunderstand the generated stimulus as an genuine stimulus for these two classes. On the contrary, there were no significant differences in the realism that participants felt between test and generated data in Carpet and Fine Foam classes. There was no correlation between the realism evaluation value and the discrimination rate of generated data.

Most scores about realism were over 60 and there was no significant difference between the scores of generated and test data in total. These results suggest that the generated vibrotactile stimuli had certain realism equivalent to the genuine stimuli. Focusing on the data for each class, the scores about realism were different between Generator Ex. and E2E Ex. This difference seems to be derived from the impression gap between texture attributes and images. Four out of ten participants answered in the post questionnaire that the impression of test image and label name were different, especially for Bamboo. Also, some participants said that it is difficult to imagine the texture surface from label name displayed in Generator Ex. These answers suggested that we should re-design the attribute axes, which we used class labels as they are in this study. For example, if we use some onomatopoeia as attribute axes, users can intuitively set and manipulate attributes and usability would be improved.

Users rated generated data significantly higher than test data for Squared Aluminum Mesh in E2E Ex. Five participants answered that they felt periodic vibrotactile stimuli like moving on the mesh as the generated stimuli, so it can be considered that the trained model has enhanced the characteristic attribute like a mesh too much.

## 5 Conclusion

In this study, we introduced the problem of vibrotactile generation based on various texture images or attributes during predefined tool-surface interaction, and solved it by adversarial training. The user study showed that users could not discriminate generated signals and genuine ones. Our approach is applicable to any case where the users touch the various surfaces in a predefined way. Thus, our study contributes to the broadening the options of vibrotactile signal preparation in such cases.